\documentclass[aps,prl,twocolumn]{revtex4}
\usepackage{amsfonts}
\usepackage{amsmath}
\usepackage{graphicx}
\usepackage{dsfont}
\usepackage{diagbox}
\usepackage{array}
\usepackage{amssymb}
\usepackage{bbm}
\usepackage{float}
\usepackage{subfigure}
\usepackage{url}
\usepackage{hyperref}
\usepackage{xcolor}
\usepackage[normalem]{ulem}

\begin{document}

\title{Gapless Coulomb state emerging from a self-dual topological
tensor-network state}
\author{Guo-Yi Zhu$^{1}$ and Guang-Ming Zhang$^{1,2}$}
\affiliation{$^{1}$State Key Laboratory of Low-Dimensional Quantum Physics and Department
of Physics, Tsinghua University, Beijing 100084, China\\
$^{2}$Collaborative Innovation Center of Quantum Matter, Beijing 100084,
China.}
\date{\today}

\begin{abstract}
In the tensor network representation, a deformed $Z_{2}$ topological ground
state wave function is proposed and its norm can be exactly mapped to the
two-dimensional solvable Ashkin-Teller (AT) model. Then the topological
(toric code) phase with anyonic excitations corresponds to the partial order
phase of the AT model, and possible topological phase transitions are
precisely determined. With the electric-magnetic self-duality, a novel
gapless Coulomb state with quasi-long-range order is obtained via a quantum
Kosterlitz-Thouless phase transition. The corresponding ground state is a
condensate of pairs of logarithmically confined electric charges and
magnetic fluxes, and the scaling behavior of various anyon correlations can
be exactly derived, revealing the effective interaction between anyons and
their condensation. Deformations away from the self-duality drive the
Coulomb state into either the gapped Higgs phase or confining phase.
\end{abstract}

\maketitle

\textit{Introduction}.-- The toric code model proposed by Kitaev\cite%
{Kitaev03ToricCode} is a prototypical model realizing the $Z_{2}$ intrinsic
topological phase of matter with anyonic excitations. It is interesting and
fundamentally important to consider the possible topological phase
transitions out of the toric code phase, because such phase transitions are
beyond the conventional Ginzburg-Landau paradigm for the symmetry breaking
phases. From the perspective of lattice gauge theory, it has been known that
there exists the Higgs/confinement transition, where the electric charge is
condensed/confined accompanied by the confinement/condensation of magnetic
flux due to electric-magnetic duality\cite%
{FradkinShenker79,Jongeward80Z2gaugeHiggs,Nayak07ToricCode,Vidal09ToricCode,StampKitaev10,Deng12ToricCode}%
. However, there is a long-standing puzzle: what is the nature of the phase
transition along the self-dual line and how the Higgs and confinement
transition lines merge into the self-dual phase transition point\cite%
{StampKitaev10,Vidal_2011}. Should there be a tricritical point, it would go
beyond the anyon condensation scenario\cite{Bais_2009}, because the electric
charge and magnetic flux are not allowed to simultaneously condense.

In this Letter, we shall resolve this puzzle and provide new insight into
the nature of this topological phase transition. Instead of solving a
Hamiltonian with tuning parameters, we propose a deformed topological
wave-function interpolating from the nontrivial to trivial phases in the
tensor network representation\cite%
{Wen10_tensor,Cirac10_peps,Verstraete15_anyonMPO}, which provides a clearer
scope into the essential physics of abelian anyonic excitations\cite%
{XieChen18,Schuch13Trsf,Verstraete15Trsf,Schuch17StringOrder,GuangMing18}.
In this scheme, the usual pure Higgs/confinement transition of the toric code%
\cite{Castelnovo08,Troyer10CQCP,Schuch13Trsf} has a special path, where the
deformed wave-function can be exactly mapped to a two-dimensional (2D)
classical Ising model. The topological phase transition is associated with
the 2D Ising phase transition, drawing the striking
topological-symmetry-breaking correspondence\cite{FengZhangXiang_2007}.

Further deformation of the toric code wave functions can span a generalized
phase diagram\cite{Verstraete15Gauging,FilterToricCode,Iqbal_2018}, where
the perturbed Higgs and confinement transitions were generically obtained by
the symmetry breaking pattern and long-range-order in the virtual space of
transfer matrix in the tensor-network formalism\cite%
{Verstraete15Trsf,Schuch17StringOrder}. But the nature of the phase
transition along the self-dual path remains elusive. By proposing a new
version of the deformed topological tensor network wave function, we prove
that the whole phase diagram can be exactly mapped to a 2D classical
isotropic Ashkin-Teller (AT) model, and the toric code phase is associated
with the partial order phase of the AT model. It is in the same spirit of
the plasma analogy for fractional quantum Hall phases\cite%
{LaughlinWaveFunction}. Such a mapping not only sheds new light on the
hidden structure of the topological phase transitions, but also enables us
to analytically pinpoint the accurate positions of the critical point and
extract the scaling behavior of the anyon correlation functions. More
importantly, we find that the toric code phase along the self-dual path
undergoes a quantum Kosterlitz-Thouless (KT) transition into the gapless
Coulomb state with quasi-long-range order\cite{Kosterlitz_1973}, which is
truly beyond the Ginzburg-Landau paradigm.

\begin{figure}[t]
\centering
\includegraphics[width=7.5cm]{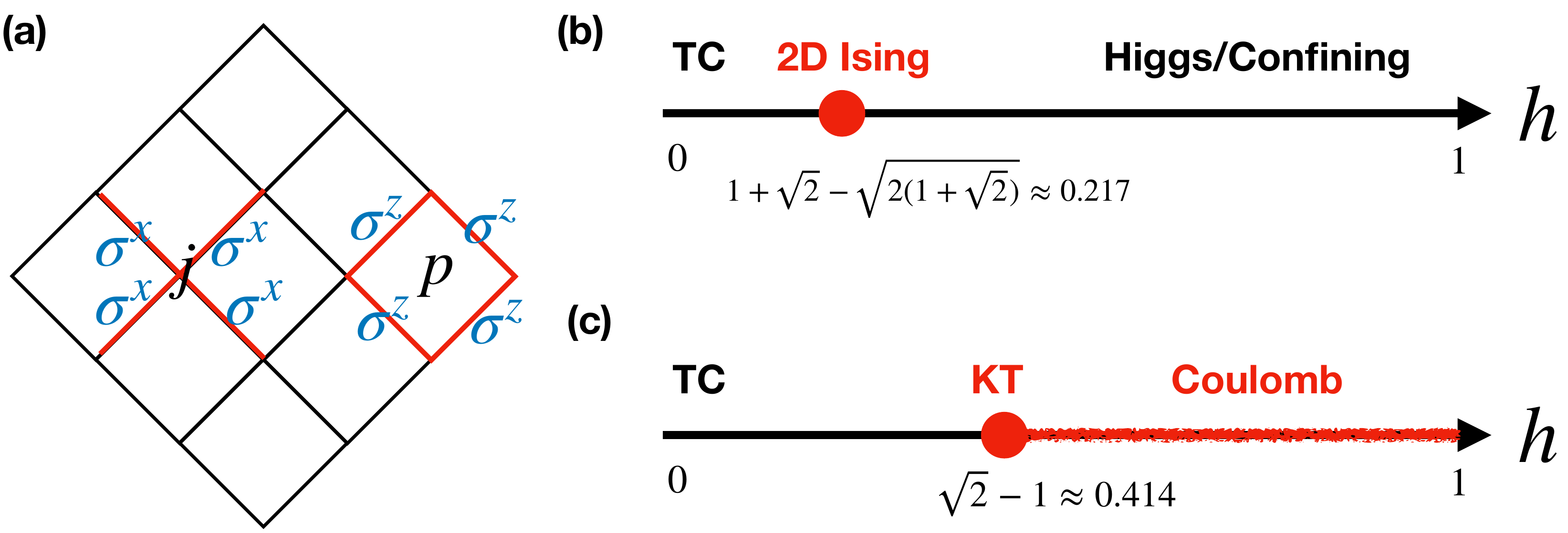}
\caption{(a) Bare Hamiltonian terms of the toric code model. (b) The 2D
Ising conformal quantum critical point encountered by the toric code phase
along the Higgs/confinement transition. (c) The 2D KT conformal quantum
phase transition of the toric code phase along the electric-magnetic
self-dual path. }
\label{Model}
\end{figure}

\textit{Wave-function deformation}.-- The toric code model on a square
lattice with periodic boundary condition is
\begin{equation}
H_{\text{TC}}=-\sum_{j}A_{j}-\sum_{p}B_{p},
\end{equation}%
where the star operator $A_{j}=\prod_{i\in \text{star}(j)}\sigma _{i,j}^{x}$
lives on the vertex $j$, and plaquette operator $B_{p}=\prod_{\langle
i,j\rangle \in \partial p}\sigma _{i,j}^{z}$ lives on the plaquette $p$
(Fig.~\ref{Model}a). The ground states are stabilized by $A_{j}=1\forall j$,
$B_{p}=1\forall p$, and $A_{j}=-1$ is associated to an electric charge
excitation while $B_{p}=-1$ to a magnetic flux excitation. The bare model is
invariant under the electric-magnetic duality:
\begin{equation}
\mathcal{D}:\ \sigma ^{x}\leftrightarrow\sigma ^{z},\ j\leftrightarrow p.
\end{equation}
If the spin up is viewed as a reference basis and the spin down as a segment
of magnetic flux tubes, the ground state is a condensate of closed magnetic
loops\cite{LevinWen05Stringnet}, or electric loops by duality. On a torus,
the four-fold ground states can be labeled by the evenness of global
noncontractible electric/magnetic loops along $x$ direction, in one-to-one
correspondence with the four anyon sectors: $1,e,m,f$. The vacuum state has
even number of both electric and magnetic noncontractible loops along $x$:
\begin{equation}
|1\rangle =\frac{1+M_{y}}{\sqrt{2}}\prod_{j}\left( 1+A_{j}\right) |\uparrow
\rangle _\sigma,
\end{equation}%
where $|\uparrow \rangle _\sigma\equiv\prod_{\langle i,j\rangle }|\uparrow
\rangle _{i,j}$ is the reference state, and $M_{y}$ is the dual Wilson line
that pumps a magnetic flux tube wrapping around the torus in $y$ direction.
It can be checked that $|1\rangle $ is invariant under electric-magnetic
duality: $\mathcal{D}|1\rangle=|1\rangle$.

In general, one can deform the toric code wave-function by filtering with
the spin polarized channel\cite{Verstraete15Gauging}:
\begin{equation}
|\psi (h,\theta )\rangle =\prod_{\langle i,j\rangle }\left[ 1+h\left( \sigma
_{i,j}^{z}\text{sin$\theta $}+\sigma _{i,j}^{x}\text{cos$\theta $}\right) %
\right] |1\rangle ,  \label{GlobalWaveFunction}
\end{equation}%
where $h\in \lbrack 0,1)$ expresses the strength of filtering and $\theta
\in \lbrack 0,\pi /2]$ is the spin angle. The limit $h\rightarrow 1$ tends
to filter out the spin polarized trivial state. When $\theta =\pi /2$, the
deformed wave-function only contains closed magnetic loop configurations and
has been well studied\cite%
{Castelnovo08,Troyer10CQCP,Troyer11CQCP,Schuch13Trsf}. The wave-function
norm is equivalent to the Ising partition function. A Rockhsar-Kivelson type
Hamiltonian is further derived by the stochastic matrix form\cite{Henley04RK}%
: $H^{z}=H_{\text{TC}}+\sum_{j}V_{j}$, with $V_{j}=\prod_{i\in \text{star}%
(j)}\left( \frac{1-h}{1+h}\right) ^{\sigma _{i,j}^{x}}$. Note $%
V_{j}\rightarrow \infty $ when $h\rightarrow 1$, which is the reason why we
require $h<1$. By electric-magnetic duality, the confining phase transition
at $\theta =0$ can be solved as well, and the phase diagram is shown in Fig.~%
\ref{Model}b. However, away from the limits $\theta =0$, $\pi /2$, it is
much less understood. Especially, $\theta =\pi /4$ corresponds to the
electric-magnetic self-dual path:
\begin{equation}
|\psi \rangle =\prod_{\langle i,j\rangle }\left( 1+h\frac{\sigma
_{i,j}^{z}+\sigma _{i,j}^{x}}{\sqrt{2}}\right) |1\rangle .
\label{WaveFunction}
\end{equation}%
The main result of this work is that, the self-dual toric code phase can
undergo a quantum KT phase transition into a gapless Coulomb state, as shown
in Fig.~\ref{Model}c.

%Rewritten into the tensor-network representation, the parent Hamiltonian can in principle be constructed to yield this tensor-network state
%as unique ground state (up to topological degeneracy)\cite{CiracParentHam}. Therefore the phase diagram can be recast to one tuned by some coupling constant.
%When $h\ll1$, to the leading order the deformed wave-function coincides with the toric code ground state perturbed by equal longitudinal and transverse magnetic field $|\psi \rangle =(1+H')|\text{0}\rangle +O\left(h^2\right)$ where $H'=\sum _{\langle i,j\rangle }h(\sigma _{i,j}^z+\sigma _{i,j}^x)/\sqrt{2}$.

\begin{figure}[t]
\centering
\includegraphics[width=7.5cm]{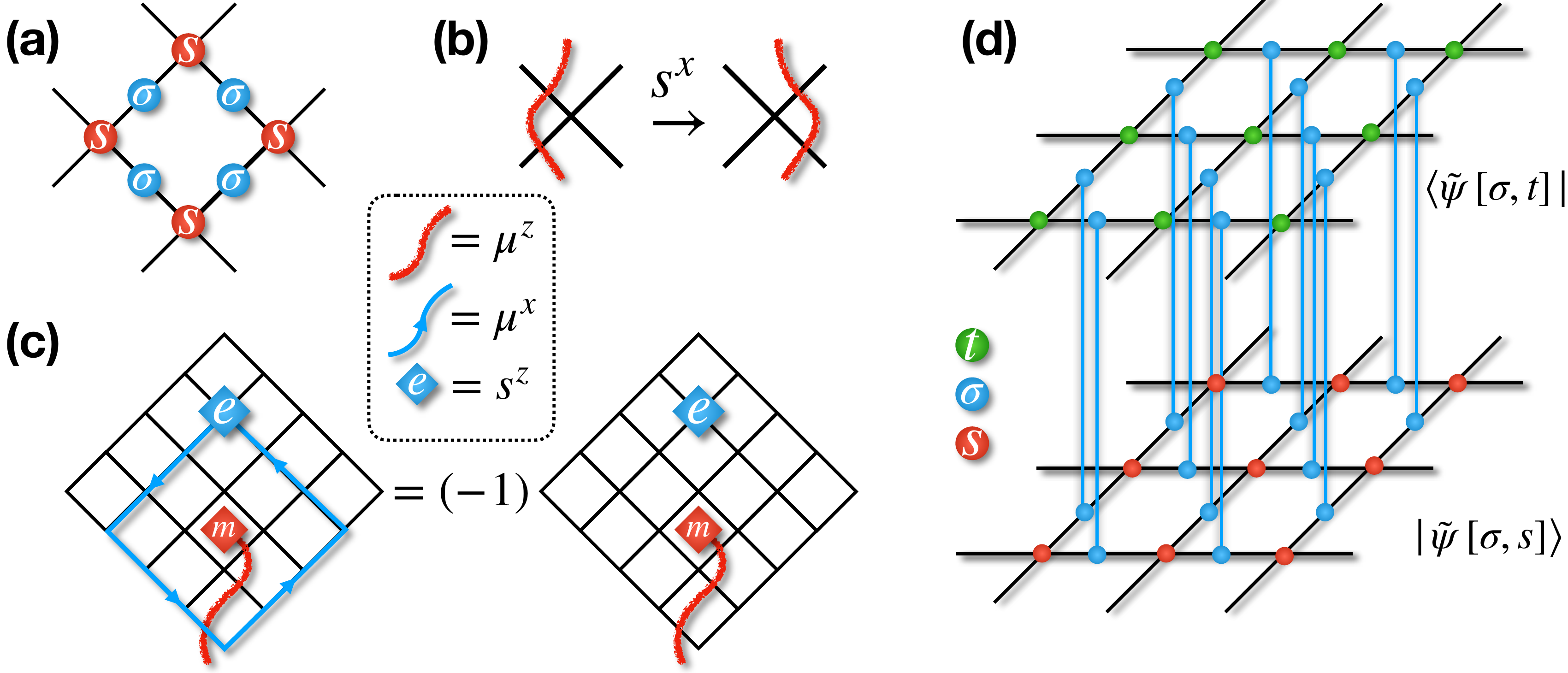}
\caption{(a) The entangled state between physical gauge spins labeled by
blue dots on the links and auxiliary matter spins labeled by red dots on the
vertices. (b) The local $Z_2$ gauge symmetry $s^x$ deforms the magnetic flux
tube generated by a sequence of $\protect\mu^z$. (c) The electric charge can
be moved by $\protect\mu^x$, and it acquires $\protect\pi$-phase across the
magnetic flux tube. (d) The norm of the quantum wave-function is composed of
double layers representing the ket and bra states, respectively. The
physical spins are contracted, while the auxiliary spins from the ket and
the bra layers are projected onto the product state $|+\rangle$ by the gauge
constraint. }
\label{Mapping}
\end{figure}

\textit{Mapping to the Ashkin-Teller model}.-- In his original paper, Kitaev
introduced the Higgs matter spins to entangle the gauge spins so as to turn
the toric code model into a gauge invariant theory\cite{Kitaev03ToricCode}.
Here we use a similar procedure to rewrite the deformed wave-function in the
extended Hilbert space with a set of auxiliary Ising spins on each vertex
restricted in the product state of $s^{x}_j$: $|+\rangle _{s}\equiv
\prod_{j}|+\rangle _{j}$. In Fig.~\ref{Mapping}a, we show the deformed
wave-function Eq.\ref{WaveFunction} expressed as an entangled state of the
physical "gauge" spins and auxiliary "matter" spins:
\begin{equation*}
|\tilde{\psi}\rangle =\prod_{\langle i,j\rangle }\left( 1+h\frac{\sigma
_{i,j}^{z}+\sigma _{i,j}^{x}}{\sqrt{2}}\right)P_{s} \frac{\left(
1+s_{i}^{z}s_{j}^{z}\sigma _{i,j}^{z}\right) }{2}|+\rangle _{s}|+\rangle
_{\sigma},
\end{equation*}%
where $P_{s}=(1+M_{y})/\sqrt{2}$ projects onto the vacuum sector. Detailed
derivation is given in Supplementary Material.

The physical subspace is subjected to the gauge constraint $s_{j}^{x}=1$,
and the auxiliary spin can be reduced by projection $|\psi \rangle =\text{ }%
_{s}\langle +|\tilde{\psi}\rangle $. The local $Z_{2}$ gauge symmetry
generated by $s_{j}^{x}$ becomes more transparent when viewed from the dual
disorder operator of $s_{j}^{z}$,
\begin{equation}
\mu _{i,j}^{x}=s_{i}^{z}s_{j}^{z},\ \ \ s_{j}^{x}=\prod_{k\in \text{star}%
(j)}\mu _{j,k}^{z}.
\end{equation}%
In Fig.~\ref{Mapping}b, we show that an arbitrary string defect of disorder
operator\cite{Vidal16Defect, Fendley16Defect, Bridgeman17_defect} $%
M_{(p,q)}\equiv \prod_{\langle i,j\rangle \in S(p,q)}\mu _{i,j}^{z}$ ending
at plaquette $p$ and $q$ is free to fluctuate under the gauge symmetry. It
can be identified as the magnetic flux tube, on whose end points live a pair
of the magnetic flux excitations $m_{p}$ and $m_{q}$ on the plaquettes.
Hence the dual Wilson line can be effectively implemented by a closed string
of disorder creator: $M_{y}=\prod_{\langle i,j\rangle }\mu _{i,j}^{z}$. The
electric charge is created by $s_{j}^{z}$, which can be moved by the action
of $\mu _{i,j}^{x}$. Since $\mu ^{x}\mu ^{z}=-\mu ^{z}\mu ^{x}$, the
electric charge acquires a $\pi $ phase whenever its trajectory crosses the
magnetic flux tube (see Fig.~\ref{Mapping}c). In this way, $m$ serves as a $%
\pi $-flux for $e$, and the semionic mutual statistics between $e$ and $m$
is thus verified. To conclude, the excitation of a pair of electric charges
and magnetic fluxes can be written as
\begin{equation}
|e_{i},e_{j}\rangle =\text{ }_{s}\langle +|s_{i}^{z}s_{j}^{z}|\tilde{\psi}%
\rangle ,\text{ }|m_{p},m_{q}\rangle =\text{ }_{s}\langle +|M_{(p,q)}|\tilde{%
\psi}\rangle .
\end{equation}

Consider the norm of the wave-function, which is viewed as a double-layer
tensor-network shown schematically in Fig.~\ref{Mapping}d. The key
observation is that if the physical spins are contracted out, the network
evaluates the partition function of the classical isotropic AT model for the
two layers of Ising auxiliary spins:
\begin{equation}
\langle \psi \lbrack \sigma ,t]|\psi \lbrack \sigma ,s]\rangle
=\sum_{\{s,t\}}P_{s}P_{t}\prod_{\langle i,j\rangle }e^{-\epsilon _{i,j}}=Z_{%
\text{AT}}[s,t],
\end{equation}%
where
\begin{eqnarray}
\epsilon _{i,j} &=&-J\left( s_{i}s_{j}+t_{i}t_{j}\right)
-J_{4}s_{i}s_{j}t_{i}t_{j}-J_{0}, \\
J &=&\frac{1}{4}\log \left( \frac{h^{2}+\sqrt{2}h+1}{h^{2}-\sqrt{2}h+1}%
\right) ,\ J_{4}=\frac{1}{4}\log \frac{1+h^{4}}{2h^{2}},  \notag
\end{eqnarray}%
and $s_{j}$ and $t_{j}$ are the quantum number of $s_{j}^{z}$ and $t_{j}^{z}$%
. The factor $P_{s}P_{t}$ implements the dual Wilson line projectors on the
ket and bra layers, equivalent to twisting the boundary condition.
%\begin{equation}
%\langle \psi [\sigma ;t]|\psi [\sigma ;s]\rangle =Z_{\text{AT}}[s, t]
%\end{equation}
Notice that the self-duality condition of the AT model $e^{-2J_{4}}=\sinh
(2J)$ is always satisfied, so that our self-dual quantum state Eq.\ref%
{WaveFunction} is exactly mapped to the self-dual isotropic AT model\cite%
{Saleur1988}. As the anyon creating operators exactly correspond to the
local spin operators or the domain wall flipping operators of the classical
AT model, the classical phase transitions serve as the faithful detections
for the topological phase transitions.

Next we briefly review the phase regime in the classical counterpart of our
quantum state. When $h<h_{c}\equiv \sqrt{2}-1$, $J_{4}>J$ and the classical
model sits in the partial order phase with $\langle s_{j}\rangle =\langle
t_{j}\rangle =0$ but $\langle s_{j}t_{j}\rangle \neq 0$, exactly
corresponding to the topological order phase of the toric code model.
However, for $h\geq h_{c}$, $J_{4}\leq J$ and the classical model enters
into the continuously varying critical phase with a fixed central charge $1$%
, described by the conformal invariant Gaussian scalar field theory\cite%
{Saleur1987,Saleur1988}: $S=\frac{R^{2}}{8\pi }\int dzd\bar{z}\partial
_{z}\phi \partial _{\bar{z}}\phi $ . The field is compactified on a circle
with the radius
\begin{equation}
R=4\sqrt{\frac{1}{\pi }\sin ^{-1}\left( \frac{1}{2}\sqrt{1+\frac{1+h^{4}}{%
2h^{2}}}\right) }.
\end{equation}%
and orbifolded $\phi =-\phi $. Thus, the transition point $h=h_{c}$ has an
enhanced symmetry $S_{4}$, corresponding to the critical point of the $q=4$
Potts model, i.e. the $Z_{2}$ orbifolding version of the KT critical point%
\cite{Dijkgraaf1988CFT}.

\textit{Anyon correlation functions}.-- With the classical phase diagram, we
can switch back to the quantum wave-function and investigate the fate of
anyons across the transition. With the electric-magnetic self-duality, the
magnetic flux is supposed to follow the same behavior as the electric
charge, so we focus on the electric charge. First, we measure the
confinement by the diagonal correlators of a pair of separated electric
charges: $\langle e_{j},e_{i}|e_{i},e_{j}\rangle $, a superposition of all
the possibly deformed spatial Wilson loops pinned by two sites $i$ and $j$.
Alternatively, it can be simply viewed as the normalization constant of the
charge excited state created from the ground state vacuum\cite%
{Verstraete15Trsf,Schuch17StringOrder}. By the quantum-classical
correspondence, we thus establish that such a diagonal anyon correlation is
equivalent to the correlation of polarization operator of the AT model,
whose asymptotic behavior is exactly known from the scaling dimension\cite%
{Kadanoff1979AT}:
\begin{eqnarray*}
\langle e_{j},e_{i}|e_{i},e_{j}\rangle &=&\left\langle
s_{j}^{z}t_{j}^{z}s_{i}^{z}t_{i}^{z}\right\rangle _{Z_{\text{AT}}} \\
&\sim &%
\begin{cases}
C^{2}\left( 1+O\left( e^{-|i-j|/\xi }\right) \right) \  & h<h_{c}, \\
|i-j|^{-\frac{2}{R^{2}}}\  & h\geq h_{c},%
\end{cases}%
\end{eqnarray*}%
where $\xi $ is the correlation length in the toric code, and $\langle \cdot
\rangle _{Z_{\text{AT}}}$ denotes the ensemble average of the classical
model.

To measure the force-law between charges, a usual way is to place a pair of
static test-charges separated in space and measure the free energy
dependence of the distance, which can be calculated by the correlations of
two Wilson lines stretching in the temporal direction of the lattice gauge
theory\cite{Svetitsky1982Confinement}. Similarly, the diagonal anyon
correlation function measures the probability amplitude of the existence of
a pair of static charges, from which we might define a dimensionless "free
energy": $\langle e_{j},e_{i}|e_{i},e_{j}\rangle \equiv e^{-F(|i-j|)}$. Then
we have
\begin{equation}
F(|i-j|)\sim
\begin{cases}
O\left( e^{-|i-j|/\xi }\right) \  & h<h_{c}, \\
\frac{2}{R^{2}}\text{log}|i-j|\  & h\geq h_{c},%
\end{cases}%
\end{equation}%
which shows the qualitative change of the spatial dependence of $F(|i-j|)$,
the hallmark of the topological phase transition. In the deformed toric code
phase with $h<h_{c}$, the result signifies a screened potential for the
deconfined charges, characteristic of a plasma phase. However, for $h>h_{c}$%
, the electric charges are weakly confined by a logarithmic potential,
characteristic of the 2D Coulomb state. The strength of the potential is
determined by the anomalous dimension of the associated correlation
function. The magnetic fluxes are likewise logarithmically confined, but the
interaction between electric charge and magnetic flux involves an additional
phase winding term, because they play the role of half vortex for each other.

Second, to consider the condensation, we measure only the pair condensation
by virtue of the off-diagonal anyon correlator, which is defined by the
overlap between the ground state and the normalized state with charges pair:
\begin{eqnarray*}
\langle \psi |e_{i},e_{j}\rangle /\left\Vert|e_{i},e_{j}\rangle\right\Vert
&=&\left\langle s_{i}^{z}s_{j}^{z}\right\rangle _{Z_{\text{AT}}}/\sqrt{%
\left\langle s_{i}^{z}s_{j}^{z}t_{i}^{z}t_{j}^{z}\right\rangle _{Z_{\text{AT}%
}}} \\
&\sim &%
\begin{cases}
O\left( e^{-|i-j|/\xi }\right) \  & h<h_{c}, \\
|i-j|^{-\left( \frac{1}{4}-\frac{1}{R^{2}}\right) }\  & h\geq h_{c},%
\end{cases}%
\end{eqnarray*}%
where $\left\Vert\cdot\right\Vert$ denotes the state norm and the scaling
dimension of $s_{j}^z$ is known as $\Delta _{s}=1/8$. There is also
qualitative change to the condensation measurement. When $h<h_{c}$, the
disorder of the classical Ising layers amounts to an exponential suppression
of the tunneling between the quantum ground state and the charge excited
state, and hence no condensation occurs. When $h>h_{c}$, the
quasi-long-range order of the classical model leads to a power-law decay of
the overlap between the ground state and charge excited pairs, indicative of
a gapless condensate of bounded charge pairs.
\begin{figure}[t]
\centering
\includegraphics[width=7.5cm]{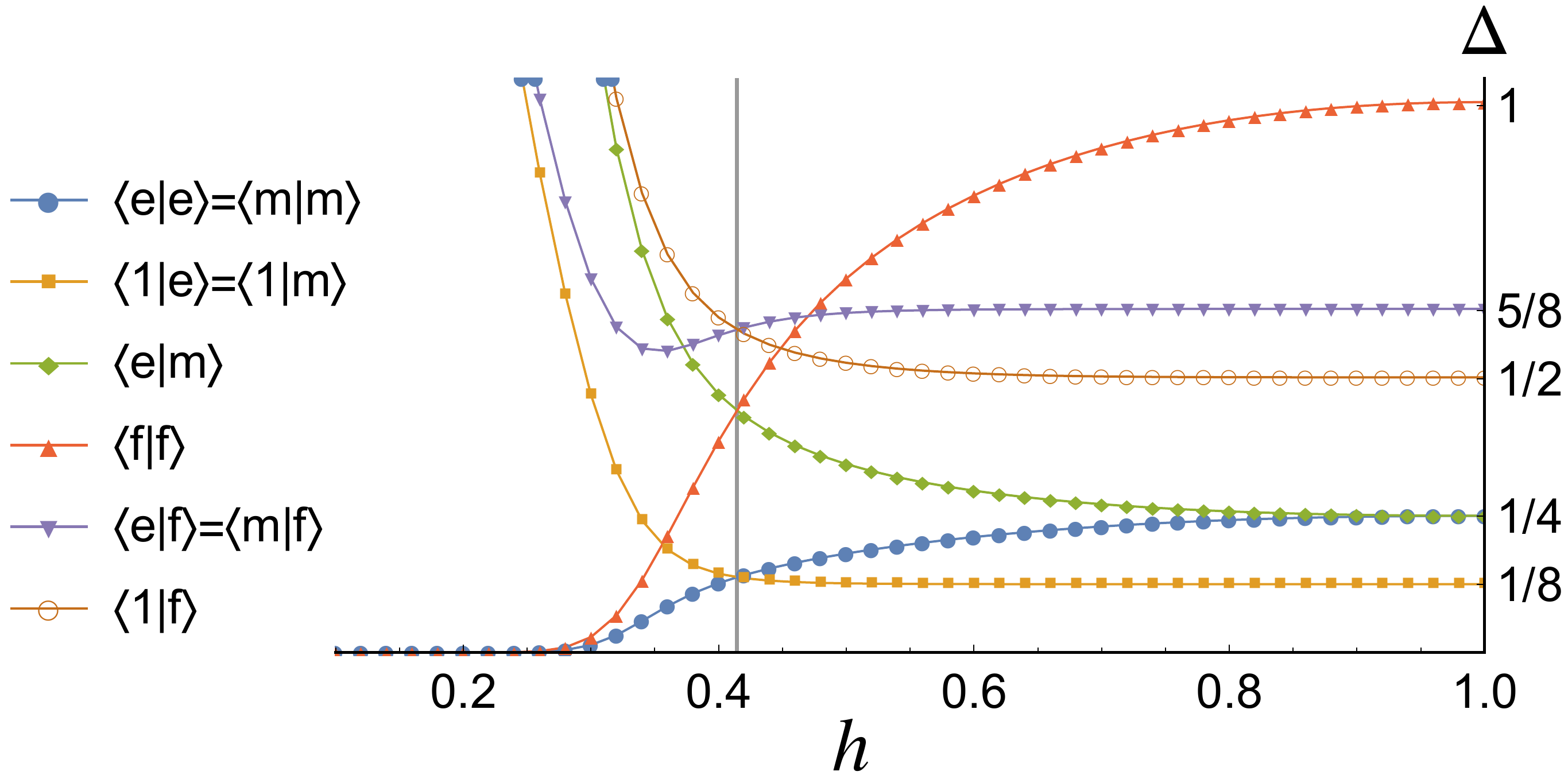}
\caption{Scaling dimensions numerically extracted from the eigenvalues of
the transfer matrix for each anyon blocks, where we set $\Delta_{\langle
1|1\rangle}=0$. The velocity parameter is estimated as $v\approx0.998$ and
the circumference of the transfer matrix is chosen as $L_y=10$. The vertical
line near $h_c=\protect\sqrt{2}-1\approx 0.414 $ denotes the KT phase
transition point.}
\label{TrsfSpec}
\end{figure}

One might ask what happens to the fermion excitations, which also are one of
the anyon types in the toric code phase. One could likewise measure $\langle
f_{j,q},f_{i,p}|f_{i,p},f_{j,q}\rangle $ for the confinement and $\langle
\psi |f_{i,p},f_{j,q}\rangle $ for the condensation. Due to the lack of the
exact scaling dimension of the string defect, we can numerically extract the
scaling dimensions of all the anyon correlations by diagonalizing the
quantum transfer matrix. As there are four topological sectors in the ket
and bra tensor layers, separately, we have 16 blocks of the transfer
operator in total, and they are labeled by $\langle \alpha |\beta \rangle $
with $\alpha ,\beta =1,e,m,f$. The leading eigenvalue for each block can be
parametrized as $\lambda _{\langle \alpha |\beta \rangle }=\lambda _{\langle
1|1\rangle }e^{-\frac{2\pi v}{L_{y}}\Delta _{\langle \alpha |\beta \rangle }}
$, where the velocity $v$ is a non-universal constant that can be fitted by
the exactly known result $\Delta _{\langle e|e\rangle }=\Delta _{s}=1/8$ at
the decoupling Ising point $h=1$. When those fermionic correlations are
expressed as
\begin{eqnarray}
\langle f_{i,p},f_{j,q}|f_{i,p},f_{j,q}\rangle &\sim &|i-j|^{-2\Delta
_{\langle f|f\rangle }},  \notag \\
\langle \psi |f_{i,p},f_{j,q}\rangle &\sim &|i-j|^{-2\Delta _{\langle
1|f\rangle }}.
\end{eqnarray}
we estimate $\Delta_{\langle f|f\rangle}\approx 4/R^2$ and $\Delta_{\langle
1|f\rangle}\approx 1/2$ slightly away from the KT transition point, as shown
in Fig.~\ref{TrsfSpec}. Then these fermions are confined by a logarithmic
potential, approximately four times that of the charges/fluxes. The full
momentum resolved spectrum of the transfer operator is included in
Supplementary Material.

\begin{figure}[t]
\centering
\includegraphics[width=7.5cm]{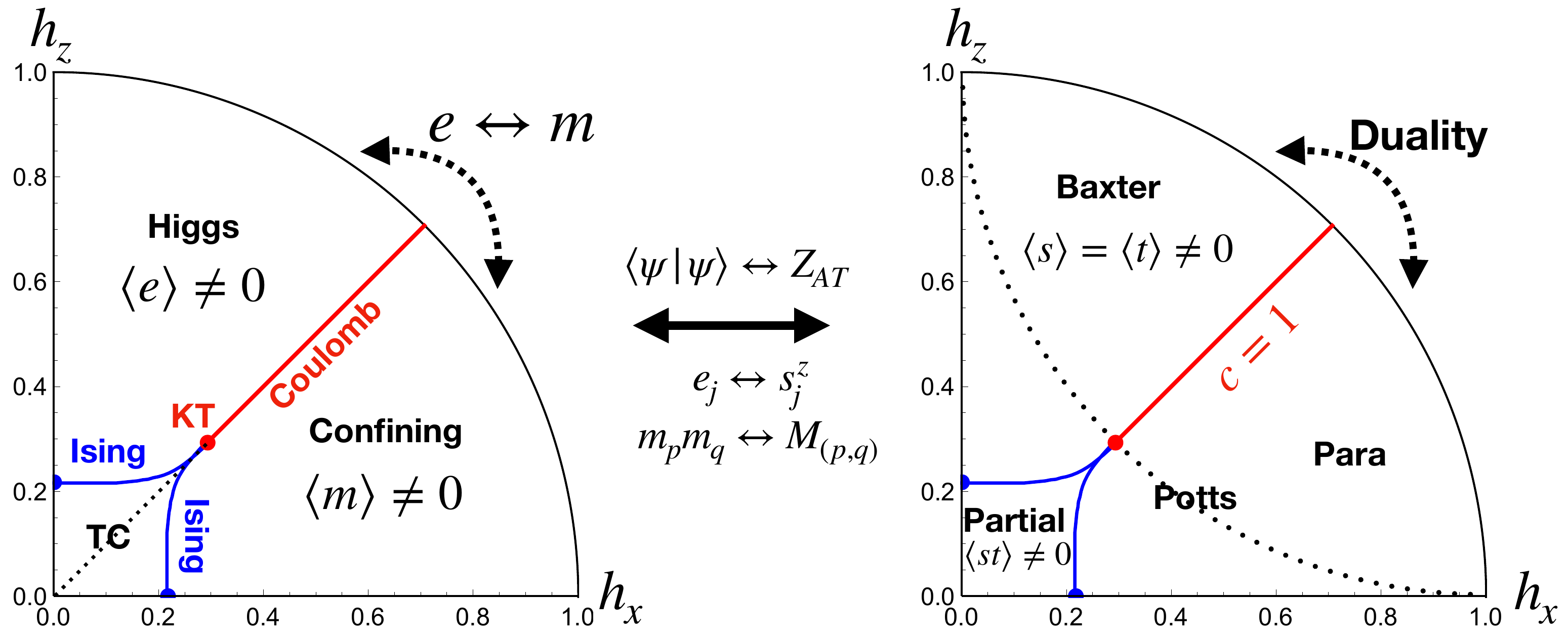}
\caption{The left panel shows the quantum phase diagram of the deformed
toric code wave-function, which can be mapped in one-to-one correspondence
to the phase diagram of the isotropic AT model on the right. The
electric-magnetic duality in the quantum wave-function coincides with the
Kramers-Wannier duality of the AT model, which transposes the phase diagram.
$h_{x}\equiv h\cos \protect\theta ,h_{z}\equiv h\sin \protect\theta $. The
blue lines are located by quantum fidelity, see supplementary material.}
\label{GlobalPhaseDiagram}
\end{figure}

\textit{Global phase diagram}.-- What is the instability of the phases in
the absence of electric-magnetic self-duality? For the global phase diagram
defined by Eq.\ref{GlobalWaveFunction}, it can be shown that the deformed
toric code wave-function along any angle $\theta $ can be exactly mapped to
the generalized isotropic AT model with
\begin{equation*}
J=\frac{1}{4}\log \frac{1+h^{2}+2h\text{sin$\theta $}}{1+h^{2}-2h\text{sin$%
\theta $}},\ J_{4}=\frac{1}{4}\log \frac{1+h^{4}+2h^{2}\text{cos2$\theta $}}{%
2h^{2}(\text{cos2$\theta $}+1)}.
\end{equation*}%
So we can draw a complete topological-classical correspondence between the
phase transitions of the deformed toric code wave-function and the AT model
(Fig.~\ref{GlobalPhaseDiagram}). The Baxter phase in the AT model\cite%
{BaxterBook} has a long range order with $\left\langle s^{z}\right\rangle
=\left\langle t^{z}\right\rangle =\left\langle s^{z}t^{z}\right\rangle \neq
0 $, corresponding to the single electric charge condensation in the quantum
state with $\left\langle \left. e_{i}\right\vert \psi \right\rangle
=\left\langle s_{i}^{z}\right\rangle /\sqrt{\left\langle
s_{i}^{z}t_{i}^{z}\right\rangle }\neq 0$, i.e. the Higgs phase. The domain
wall costs an energy linearly proportional to its length, signifying the
confinement of magnetic fluxes. In the disorder paramagnetic phase of the AT
model, $\left\langle s^{z}\right\rangle =\left\langle t^{z}\right\rangle
=\left\langle s^{z}t^{z}\right\rangle =0$ and the correlation function
decays exponentially with a correlation length, and a linear confining
potential is resulted $F(|i-j|)=-\log \langle e_{j},e_{i}|e_{i},e_{j}\rangle
\sim |i-j|/\xi $ for the electric charges, corresponding to the confining
phase.

The Coulomb state corresponds to the self-dual continuously varying
criticality of the AT model, which further bifurcates at the critical point
of the $q=4$ Potts model into two 2D Ising transition lines accounting for
the Higgs and confining transitions, respectively. When the self-duality is
absent, the gapless Coulomb state falls immediately into the confining phase
or Higgs phase. In the former phase, the magnetic flux condenses and
develops a linearly confining potential for the electric charges, while in
the latter phase the electric charges get deconfined and condensed. On the
critical line, the electric charge is weakly confined and there emerges a
U(1) symmetry. In this sense, such a transition shares certain similarities
with the deconfined quantum criticality\cite{Senthil}. It is expected that
the effective theory comprises a noncompact U(1) gauge field coupled to the
massless bosonic charge, and the quantum KT critical point characterizes the
charge-2e Higgs transition.

\textit{Conclusion and Outlook}.- -- We have elucidated the topological
phase transitions out of the toric code phase along the electric-magnetic
self-duality. The novel topological phase transition is beyond the
conventional Ginzburg-Landau paradigm, exhibiting the time-independent
conformal symmetry\cite{Ardonne04CQCP} which usually occurs in the
Rokhsar-Kivelson type models\cite{RK88}. The obtained tensor-network wave
functions can serve as a variational ansatz for the deformed model
Hamiltonians. As the bond dimension of local tensors increases, the more
generic phases can be explored\cite{Lewenston_2014,Troyer11CQCP}. We expect
that the critical line of the Coulomb state becomes a first-order phase
transition line, and the quantum KT transition could flow to its Lorentz
invariant analog, namely, a 3D XY critical point characterizing the
charge-2e Higgs transition between the $Z_2$ deconfined phase and the $U(1)$
confined phase\cite{FradkinShenker79,Sondhi04SC}. Therefore, our study
provides a useful method to tackle the global phase diagram of the generic
toric code Hamiltonian\cite{StampKitaev10}. Meanwhile, it can also be
generalized to investigate the $Z_n$ gauge deconfined phases\cite%
{Wen09StringnetTensornetwork,Vidal09StringnetTensornetwork} and the
nonabelian phase\cite{Kitaev06Anyon,Kawashima_2019}.

\textit{Acknowledgment}.- The authors would like to thank Wen-Tao Xu and
Shao-Kai Jian for their stimulating discussions and acknowledge the support
by the National Key Research and Development Program of MOST of China
(2017YFA0302902)).

\newpage
\begin{widetext}

\section{Supplementary Material for "Gapless Coulomb phase emerged from a self-dual topological
tensor-network state"}

\subsection{Introduction of the auxiliary matter spins}

In the toric code model, the electric charge or magnetic flux is conserved modulo 2 even under perturbations as long as the phase transition does not occur. This implies a hidden $Z_2\times Z_2$ symmetry. To explain the origin of such a symmetry, Kitaev extended the Hilbert space and introduced the Higgs matter field on vertex/plaquette, which is yet subjected to a local constraint\cite{Kitaev03ToricCode}. The artificial constraint actually is lifted to a $Z_2$ gauge symmetry and has rather physical meaning. Especially, after a unitary transform for the extended Hilbert space, the gauge symmetry is in a more familiar look as in the conventional lattice gauge theory\cite{FradkinShenker79}. In dealing with the wave-function, one could do a similar procedure. For the sake of clearness we show how to introduce the matter spin to entangle with the physical gauge spin in the simplest plain ground state wave-function without deformation or global projection: $|\psi_0\rangle =\prod _j\frac{1+A_j}{\sqrt{2}}|\uparrow \rangle_{\sigma }$.

First we extend the Hilbert space and introduce the auxiliary matter spin in the $x$ polarized state $|+\rangle_s$ on the vertex:
\begin{equation}
|\psi_0\rangle \rightarrow |\tilde{\psi_0}\rangle\equiv  P_G    |\psi_0\rangle\otimes |+\rangle_s,
\end{equation}
where the projector $P_G=\prod_j \frac{1+s_j^x}{2}$ is to enforce the local gauge constraint $s_j^x=1$ in the extended state. This might seem trivial at first glance, but if the gauge constraint is strictly enforced, within the gauge invariant physical space one can derive that
\begin{equation}
\begin{split}
|\tilde{\psi}_0\rangle &=P_G \prod _j\frac{1+A_j}{\sqrt{2}}|+\rangle_s |\uparrow\rangle_{\sigma } \\
&=P_G \prod _j\frac{1+s_j^xA_j}{\sqrt{2}}|+\rangle_s |\uparrow\rangle_{\sigma } \\
&=P_G\prod _j\left(1+s_j^xA_j\right)\frac{1+s_j^x}{2}|\uparrow \rangle _s|\uparrow \rangle_{\sigma }\\
&=P_G\prod _j\left(1+s_j^xA_j\right)|\uparrow \rangle_s |\uparrow \rangle_{\sigma } \\
&= P_G\sum _{\{s, \sigma \}}\prod _{\langle i,j\rangle }\frac{1+s_j\sigma _{i,j}s_i}{2}|s\rangle |\sigma \rangle \\
&= P_G\prod _{\langle i,j\rangle }\frac{1+s_j^z\sigma _{i,j}^zs_i^z}{2}|+\rangle_s |+\rangle_{\sigma }.
\end{split}
\end{equation}
We can see that even without a unitary transform to change the looking of the gauge symmetry, one can rewrite the wave-function into an entangled state between the physical gauge spins and auxiliary spins, by virtue of the gauge constraint. To reduce the redundant matter spins in the wave-function, one can simply project the extended state onto the gauge invariant subspace:
\begin{equation}
\begin{split}
|\psi_0\rangle&=\text{}_s\langle+|\tilde{\psi}_0\rangle\\
&= \text{}_s\langle+|\prod _{\langle i,j\rangle }\frac{1+s_j^z\sigma _{i,j}^zs_i^z}{2}|+\rangle_s |+\rangle_{\sigma }\\
&= \sum_{\{s\}}\prod _{\langle i,j\rangle }\frac{1+s_js_i\sigma _{i,j}^z}{2} |+\rangle_{\sigma }.
\end{split}
\end{equation}
The introduction of auxiliary matter spins is robust against the physical deformation, and therefore we have the extended deformed wave-function:
\begin{equation}
|\tilde{\psi}(h,\theta) \rangle = P_G \prod _{\langle i,j\rangle }\left(1+h \left( \sigma _{i,j}^x\text{cos$\theta $}+ \sigma _{i,j}^z\text{sin$\theta $}\right)\right) \frac{1+M_y}{\sqrt{2}}\frac{1+s_j^z\sigma
_{i,j}^zs_i^z}{2}|+\rangle_s |+\rangle_{\sigma},
\end{equation}
whose projection onto the gauge invariant subspace reduces the auxiliary degrees of freedom:
\begin{equation}
|\psi(h,\theta) \rangle =\text{}_s\langle+|\tilde{\psi}(h,\theta) \rangle.
\end{equation}
A brief remark: one can see that actually in this way the wave-function turns to the tensor-network formalism, where the quantum numbers of the auxiliary matter spins play the role of the virtual bond dummy variables. That is the reason why the symmetry breaking pattern in the virtual space of the transfer matrix\cite{Schuch13Trsf,Verstraete15Trsf} can correspond to the Higgs transition, where the gauge symmetry is spontaneously broken due to the matter field.

\subsection{Derivation of the Boltzman weight}

When restricted to the given wave-function, the auxiliary spin $s_j$ plays the role of a dummy variable in generating the wave-function $|\psi\rangle$. The auxiliary spin that generates $\langle\psi|$, on the other hand, has nothing to do with and should be independent from the $s_j$, which we might label as $t_j$. Therefore, we have two layers of Ising spins in representing a generic correlation function out of the ground state, which is consistent with the tensor-network formalism. By contracting the physical degree of freedom in the wave-function,
\begin{equation}
|\psi (h,\theta )\rangle =\sum _{\{s\}, \{\sigma \}}P_s\prod _{\langle i,j\rangle }\left(\left(1+h \sin\theta \sigma _{i,j}\right)\frac{1+s_is_j\sigma
_{i,j}}{2}+h \cos\theta \frac{1-s_is_j\sigma _{i,j}}{2}\right)|\sigma _{i,j}\rangle ,
\end{equation}
where $\sigma $ is the quantum number of $\sigma ^z$, one can derive the norm as the partition function of $\{s,t\}$:
\begin{equation}
\langle \psi |\psi \rangle =\sum _{\{s,t \}}P_sP_t\prod _{\langle i,j\rangle }\omega _{i,j},
\end{equation}
where the local "Boltzman weight" is
\begin{equation}
\omega _{i,j}=\left(\left(1+h \sin\theta s_is_j\right)^2+h^2\cos ^2\theta \right)\frac{1+s_is_j\sigma _i\sigma _j}{2}+2h \cos\theta\frac{1-s_is_j\sigma _i\sigma _j}{2}\equiv e^{-\epsilon _{i,j}}.
\end{equation}

As there are two independent parameters $(h,\theta )$, taking the global normalization constant into account, one has three parameters. Therefore, three energy levels for the 16 local configurations consist of 4 spins $\left(s_i,s_j,t_i,t_j\right)$, which can be parametrized by three
parameters in the energy form $\epsilon _{i,j}=-J \left(s_is_j+t_it_j\right)-J_4s_is_jt_it_j-J_0$. The 2-spin, 4-spin interactions and the constant terms are given as
\begin{equation}
\begin{split}
&J=\frac{1}{4}\log \frac{1+h^{2}+2h\text{sin$\theta $}}{1+h^{2}-2h\text{sin$
\theta $}},\\
&J_{4}=\frac{1}{4}\log \frac{1+h^{4}+2h^{2}\text{cos2$\theta $}}{
2h^{2}(\text{cos2$\theta $}+1)},\\
&J_0=\frac{1}{4}\log \left(\left(1+h^4+2h^2\text{cos2$\theta $}\right)2h^2(\text{cos2$\theta $}+1)\right).
\end{split}
\end{equation}
This is nothing but the classical isotropic Ashkin-Teller(AT) model, where two layers of Ising model are coupled by a four-spin interaction, as one of the natural extension lists of the well-known Ising model. It has been shown to be mapped to the staggered eight vertex model\cite{BaxterBook}. Especially, along the self-dual line with $e^{-2J_{4}}=\sinh(2J)$, it can be mapped to a line in the phase diagram of homogeneous eight vertex model, as well as dually mapped to the six vertex model, where there is a continuously varying U(1) critical line separated from the disorder phase by the Kosterlitz-Thouless transition. Therefore, the AT model is one of the central models in the statistical mechanics whose criticality has been exactly solved. Notice that despite their similarities, distinct from the eight vertex model, the AT model has an exceptional partial order phase, whose phase boundary is the Ising transition lines bifurcated from the $q=4$ Potts critical point. In our parametrization, the duality of the AT model is as simple as a flipping of the $h_x\leftrightarrow h_z$ or $\theta\leftrightarrow \pi/2-\theta$. And the self-dual line lies on the $\theta=\pi/4$.

\subsection{The correspondences of local operators}

To extract the local operator correspondence between the quantum physical space and the classical auxiliary space, one has to insert the physical operator onto the physical level and contract out the physical spin, leaving a factor represented by the auxiliary spin $\omega'[s,t;\sigma^\alpha]$, where $\alpha=0,x,y,z$. Then by subtracting the Boltzmann weight, one arrives at the corresponding classical operator $\tilde{\sigma}^\alpha\equiv \omega'[s,t;\sigma^\alpha]/\omega[s,t]$, such that
\begin{equation}
\langle\psi|\sigma_{i,j}^\alpha \sigma_{k,l}^\beta\cdots|\psi\rangle=\langle\tilde{\sigma}_{i,j}^\alpha \tilde{\sigma}_{k,l}^\beta\cdots\rangle_{Z_{\text{AT}}}.
\end{equation}
This procedure could be performed generically, but here we only show the result along the self-dual line:
\begin{equation}
\begin{split}
\tilde{\sigma}_{i,j}^z&=\frac{1}{2\sqrt{2}(1+h^4)}\left(  h(1+h^2)^2+\sqrt{2}(1-h^2)(s_is_j+t_it_j)-h(1-h^2)^2 s_is_jt_it_j    \right),\\
\tilde{\sigma}_{i,j}^x&=\frac{1}{2\sqrt{2}(1+h^4)}\left(\frac{1}{h}(1+h^2)^2-\sqrt{2}h^2(1-h^2)(s_is_j+t_it_j)-\frac{1}{h}(1-h^2)^2 s_is_jt_it_j   \right),\\
i\tilde{\sigma}_{i,j}^y&=\frac{1-h^2}{2\sqrt{2}h}\left(s_is_j-t_it_j\right).
\end{split}
\end{equation}

If one is careful enough to investigate into the expression above, one could be alert to find that at some limit points the mapping between some of the local operators could break down. For example, the limit $h=1$ has $\tilde{\sigma}^{z}=\tilde{\sigma}^{x}=1/\sqrt{2},\ \tilde{\sigma}^y=0$, so that the local operators in the quantum state is blinded to any singularity behaviour at the classical counterpart, which is in the decoupled Ising critical point $(\text{Ising})^2$. Indeed, at the limit $h=1$ the quantum state is just a trivial product state, although the classical model is in the decoupled Ising critical point. From another angle it is reasonable for the mapping to break down at the particular decoupled Ising point, because the ket and bra layer are never supposed to be completely "decoupled". In this sense, $h=1$ is a very special limit point that cannot represent the physics in $h<1$, justifying our choosing the phase regime $h\in[0,1)$. Another hint that justifies excluding the $h=1$ limit is some coupling constants in the RK Hamiltonian blow up towards the limit $h\rightarrow1$.

\subsection{Quantum fidelity for the phase transitions}

With the exact ground state in hand, we could numerically probe the phase diagram by using quantum fidelity straightforwardly\cite{Zanardi07Fidelity,Verstraete15Gauging}. We measure the quantum fidelity $F_h$ when h is perturbed with $\theta$ being fixed, and the quantum fidelity $F_\theta$ when h is fixed but $\theta$ is varied:
\begin{equation}
\begin{split}
&F_h\equiv\langle \psi(h,\theta)|\psi(h+\delta h, \theta)\rangle= e^{-g_h\delta h^2 N},\\
&F_\theta\equiv\langle \psi(h,\theta)|\psi(h, \theta+\delta\theta)\rangle= e^{-g_\theta h^2\delta \theta^2 N},
\end{split}
\end{equation}
from which we can extract the quantum fidelity metric along the radial (azimuthal) direction $g_h(\theta)$ with $N$ being the total number of lattice sites. The numerical calculation is shown in Fig.~\ref{Fidelity}. Notice that, without the support of the analytical results, it would be tough to pinpoint the accurate position of the critical point of the 4-state Potts model by merely numerical calculation, which suffers from the logarithmic finite size effect, characteristic of KT transition\cite{Vekua15FidelityKT}.
\begin{figure}[htbp]
   \centering
   \includegraphics[width=12cm]{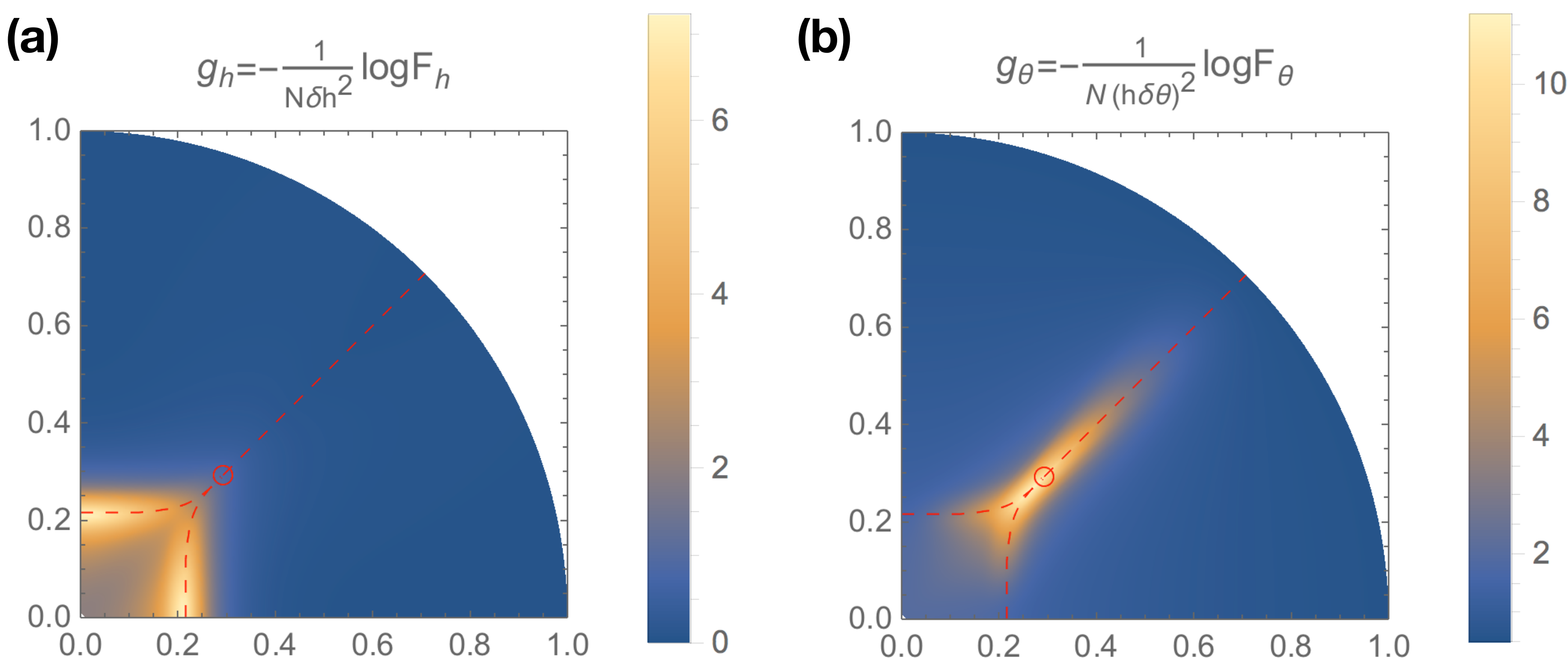}
   \caption{Quantum fidelity metric. (a) $\theta$ is fixed, $h$ is varied. (b) $\theta$ is varied, $h$ is fixed. The red dashed line marks the location of metric peak, i.e. the phase transition lines. The red open circle denotes the exact location of the Potts point to guide the vision. In the calculation of wave-function overlap, we take a $(L_y=6)\times (L_x=100)$ tensor-network , where the local tensors are located on half of the plaquette centers.}
   \label{Fidelity}
\end{figure}

\subsection{Full spectrum of the transfer matrix at the critical points}

We calculate a complete spectrum of the transfer matrix at the phase transition point $h=\sqrt{2}-1$, resolved by momentum in Fig.~\ref{TrsfSpecPotts}a. The transfer matrix has 10 independent blocks corresponding to the four anyon sectors in bra and the four anyon sectors in ket. The lowest excited levels corresponding to the anyon block $\langle e|e \rangle$ and $\langle 1|e \rangle$ have exact scaling dimension 1/8 as expected analytically, and the corresponding conformal spins (rescaled momentum $kL_y/2\pi$) are 0. The lowest level in the block $\langle f|f\rangle$ is close to (0,1/2), parametrized by the Gaussian theory $\Delta=4/R^2$, and the lowest level in block $\langle f|1\rangle$ is close to (1/2,1/2). For comparison we also give the spectrum at the Ising decoupled limit $h=1$ (Fig.~\ref{TrsfSpecPotts}b). Although the limit point $h=1$ is excluded from our phase diagram, by comparing the spectrum between the transition point $h=\sqrt{2}-1$ and that of the $h=1$ point, one can get a glimpse of the evolution of the spectrum along increasing $h$. Except some levels due to the $Z_2$ orbifolding such as $\Delta=1/8$, most of the primary levels can be rewritten in the Gaussian language labeled by the spin wave index n of operator $e^{i n\phi}$ and vorticity index p of the scalar field: $(n,p)$\cite{Batchelor1988}. The corresponding scaling dimension and conformal spin are parametrized by the radius
\begin{equation}
\Delta=\frac{n^2}{R^2}+\frac{p^2R^2}{4},\ s=n p.
\end{equation}
For the Potts point, $R=2\sqrt{2}$, and $R=2$ for the decoupled Ising point. Note that we avoid the more common notation with e and m to denote the spin wave index and vortex index, so as to avoid confusion between the charge and vortex in the language of scalar field theory and that of the toric code.

\begin{figure}[htbp]
   \centering
   \includegraphics[width=12cm]{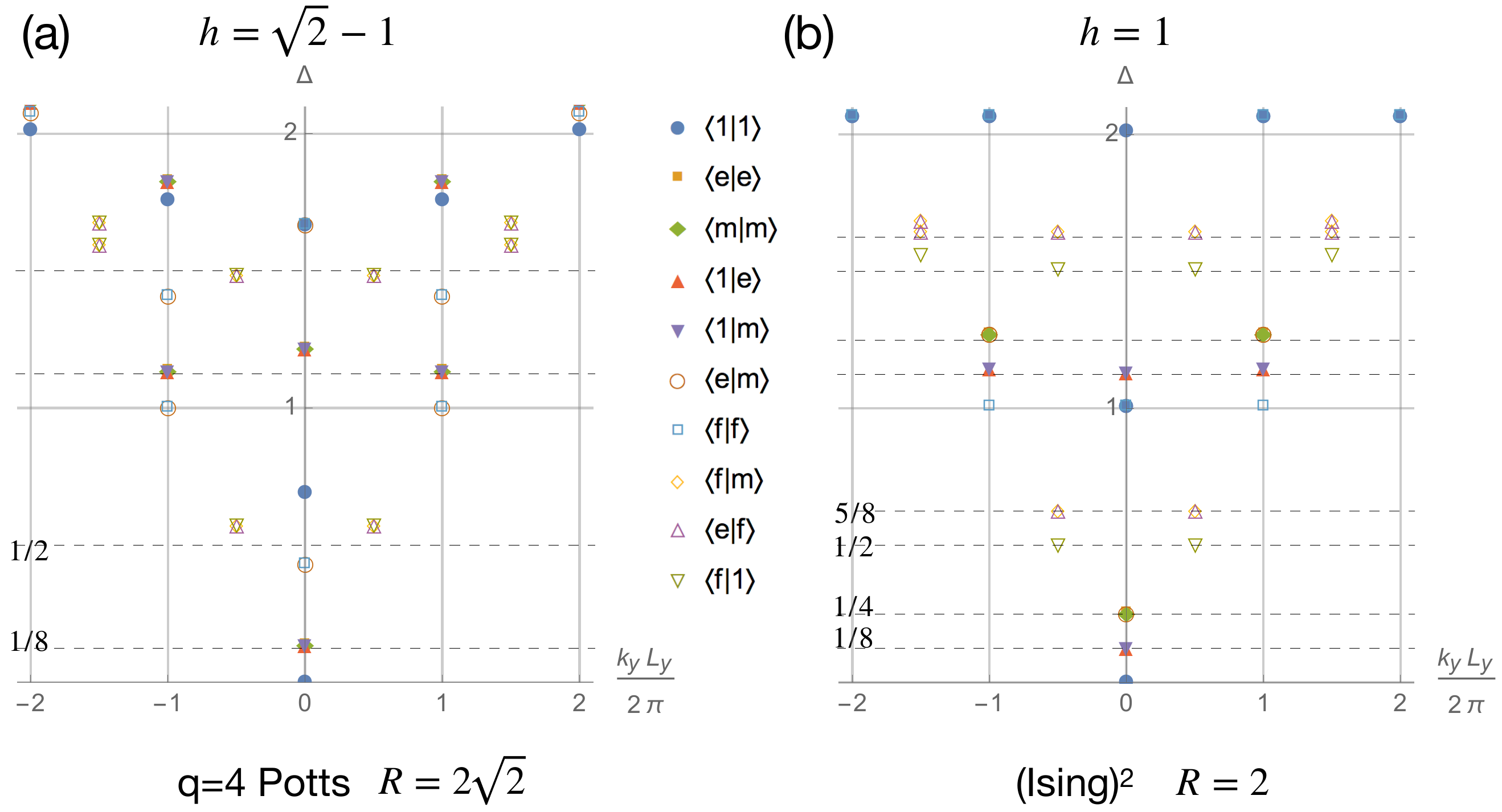}
   \caption{(a) Exact diagonalization for the transfer matrix at the transition point $h=\sqrt{2}-1,\theta=\pi/4$, resolved by momentum along the y direction. The eigenvalue $\lambda _{\langle\alpha|\beta\rangle}=\lambda _{\langle 1|1\rangle}e^{-\frac{2\pi v}{L_y}\Delta _{\langle\alpha|\beta\rangle}}$. $v\approx 1.028$ and $L_y=10$. The figure shows the corresponding scaling dimension $\Delta$ and scaled momentum for the levels in each anyon sectors. The dashed line marks the position of particular scaling dimension $\Delta=1/8,1/2 $ and their descendants. In this numerical calculation the finite size effect is very strong. (b) Exact diagonalization for the transfer matrix at the Ising decoupled limit $h=1,\theta=\pi/4$, resolved by momentum along the y direction. $v\approx 0.998$ and $L_y=10$. The dashed line marks the position of particular scaling dimension $\Delta=1/8,1/4,1/2,5/8$ and their descendants.}
   \label{TrsfSpecPotts}
\end{figure}
\end{widetext}

\bibliography{GuoYiReference}

\end{document}